# A Compact Rejection Filter Based on 2.5-D Spoof Surface Plasmon Polaritons and Folded Spilt-Ring Resonators

Hong-Bin Zhu, Lei Ji, Xiao-Chun Li, *Senior Member, IEEE*, and Jun-Fa Mao, *Fellow, IEEE*

Key Laboratory of Ministry of Education of Design and Electromagnetic Compatibility of High Speed
*Electronic Systems, Shanghai Jiao Tong University*
Corresponding author: Xiao-Chun Li, lixc@sjtu.edu.cn

*Abstract* —In this paper, a compact rejection filter is proposed based on 2.5-D spoof surface plasmon polariton (SSPP) structure and folded spilt-ring resonators (SRRs). By embedding the folded SRRs into periodic grooves of the 2.5-D SSPP transmission line, a rejection band is created in the ultra-wide passband, resulting in pass-rejection-pass frequency characteristics. The rejection band and the cut-off frequency of the filter can be adjusted independently by changing the geometrical parameters of the 2.5-D SSPP rejection unit. Because of the compact folding structure, the lateral size is reduced by 54% compared with the conventional 2-D SSPP rejection filter.

*Index Terms* — dispersion feature, folded spilt-ring resonator, rejection filter, spoof surface plasmon polariton

## I. INTRODUCTION

Spoof surface plasmon polariton (SSPP) is the structure mimicking transmission characteristics of SPP at the interface between metal and medium in the microwave frequency band using artificial periodic array [1][2]. Due to the advantages of SSPPs, such as the field confinement, adjustable dispersion characteristics and the slow wave effect, they have been applied in the design of filters [3-8]. The dispersion characteristics can be effectively controlled in the target frequency band by adjusting the geometrical parameters of the SSPP unit. Therefore, compared with conventional microwave devices, the filters based on SSPP structure have the advantages of miniaturization and controllable performance.

Rejection filters can create a rejection band in the ultra-wide passband, presenting the frequency characteristics of pass-rejection-pass, which can be used to suppress interference signals. As frequency resources become increasingly scarce, rejection filters are of great significance in today's ultra-wideband systems. In 2013, resonators were loaded on both sides of the double-groove SSPP to reject the transmission of electromagnetic waves at some frequency points in the passband [5]. However, the loaded resonators increased the lateral size of the filter. In 2016, a single rejection filter was realized by etching complementary split-ring resonators (SRRs) on double-groove SSPP TLs, which reduced the lateral width [6]. In 2017 and 2019, different forms of rectangular SRRs were loaded in the periodic grooves of the single-groove SSPP TLs, which also realized single rejection filters without increasing the lateral size [7][8]. However, the size of the resonant structure is restricted by the lateral size of the SSPP structure, which limits the rejection band adjustment range.

In this paper, a compact rejection filter based on the 2.5-D SSPP [4] and folded SRRs is proposed. The rejection band and the cut-off frequency of the filter can be adjusted independently by properly changing the geometrical parameters of the 2.5-D SSPP rejection unit. Compared with conventional 2-D SSPP rejection filter, the proposed 2.5-D SSPP filter uses compact folding structure, which can reduce the lateral size by 54%. Experimental results show that the two passbands are from DC to 5.98 GHz and 6.38 to 9.15 GHz with the insertion loss (IL) less than 2.0 dB. The rejection band locates around 6.13 GHz with its 3-dB bandwidth from 5.92 to 6.32 GHz, and the maximum rejection depth is up to -40 dB.

## II. CONFIGURATION OF THE PROPOSED STRUCTURE

The configuration of the proposed rejection filter is shown in Fig.1, which consists of two parts: transition part (Region I), and transmission part (Region II). In the transmission part, five 2.5-D SSPP rejection units are loaded, which are formed by embedding the folded SRRs into the grooves of the 2.5-D SSPP units. The proposed filter consists of two substrate layers sandwiched in three metal layers, and $h_s$ is the thickness of the total substrate. In each 2.5-D SSPP unit, a sub-wavelength groove is etched onto the top metal layer and the top layer is connected to the middle layer by metalized blind holes. The geometrical parameter $dd_1$ is the diameter of the metallized blind holes, $H$ and $p$ are the width and length of the periodic unit, and $w_a$, $d_t$ and $d_b$ are the width and depth of the corrugated structures on the top and middle metal layers, respectively.

The folded SRR consists of a U-shaped open resonant ring on the top metal layer and coupling lines bending inward on the middle metal layer. $w_b$ and $l_t$ are the width and length of the U-shaped open resonant ring on the top metal layer, $w$ is the width of the line, $l_b$ and $g$ are is the length and the gap width of the coupling lines on the middle metal layer, respectively, and $dd_2$ is the diameter of the metallized blind holes. Compared with conventional 2-D SSPP rejection unit, the proposed 2.5-D SSPP rejection unit has the extra middle metal layer, which can be regarded as folding the top layer through metal blind holes.

In the transition part, tapering 2.5-D SSPP units with gradient length are designed at both ends as the microstrip-to-SSPP transitions with good impedance and mode matching. The parameter $w_s$ is the width of microstrip line, and $l_m$ is the length of the transitions.

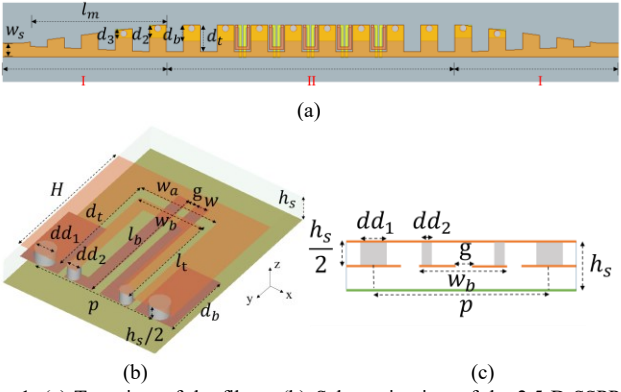

Fig. 1. (a) Top view of the filter. (b) Schematic view of the 2.5-D SSPP rejection unit. (c) x-z cross section of the 2.5-D SSPP rejection unit.

## III. SIMULATION AND EXPERIMENTAL RESULTS

The proposed structure is fabricated using the PCB process with its geometrical parameters shown in Table I. The material of the substrate layer is Rogers 4350 substrate with relative permittivity $\varepsilon_r$=3.66 and loss tangent $tan\delta$=0.004. Full-wave simulations and experimental measurement are performed.

### A. Dispersion analysis of the 2.5-D SSPP rejection unit

In order to analyze the dispersion characteristics of the proposed structure, dispersion curves of the 2.5-D SSPP rejection unit with the folded SRR integrated into the groove are shown in Fig.2(a). In the figure, the solid lines represent the dispersion curves of the first two modes (mode 0 and mode 1) of the 2.5-D SSPP rejection unit, while the dotted line is the dispersion curve of the 2.5-D SSPP unit with the same size for comparison. Light line represents the dispersion characteristic of electromagnetic wave propagating in vacuum. Here, $\beta$ is the propagation constant along the propagation x-direction.

According to the SSPP theory [2], the transmission passband of SSPP is located on the right side of the light line and the area below the cut-off frequency of the dispersion curve. As can be seen from Fig.2(a), the dispersion curve of mode 0 of the 2.5-D SSPP rejection unit is completely on the right of the light line, so the area below this line is the transmission passband of mode 0. While, the starting frequency of the dispersion curve of mode 1 is higher than that of the cut-off frequencies of mode 0. Therefore, a forbidden band marked in grey is formed between the passbands of mode 0 and mode 1, presenting the frequency characteristics of pass-rejection-pass. The cut-off frequency of the mode 1 is consistent with that of the 2.5-D SSPP unit.

### B. Analysis of the transmission features of the rejection filter

Based on the dispersion analysis, the proposed 2.5-D SSPP rejection unit can be applied in designing a rejection filter. The simulated transmission features of the proposed rejection filter

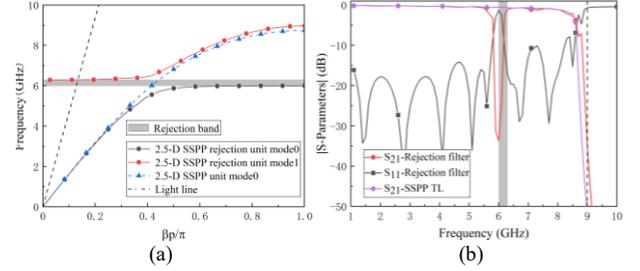

Fig. 2. (a) Dispersion curves of the 2.5-D SSPP rejection unit, the 2.5-D SSPP unit with the same size and the light line. (b) |S-parameters| of the proposed rejection filter and the 2.5-D SSPP TL.

TABLE I
DIMENSIONS OF THE PROPOSED STRUCTURE

| Para. | Dim. (mm) | Para. | Dim. (mm) | Para. | Dim. (mm) |
|---|---|---|---|---|---|
| $h_s$ | 0.508 | $d_3$ | 0.8 | $w_b$ | 1.4 |
| $w_s$ | 1.3 | $H$ | 3 | $l_t$ | 2.4 |
| $l_m$ | 13.6 | $dd_1$ | 0.6 | $l_b$ | 3 |
| $d_t$ | 2.5 | $dd_2$ | 0.2 | $g$ | 0.1 |
| $d_b$ | 1.5 | $p$ | 3.2 | $w$ | 0.25 |
| $d_2$ | 1.2 | $w_a$ | 1.6 | | |

with five folded SRRs, as well as the 2.5-D SSPP TL for comparison, are depicted in Fig.2(b).

It can be seen that the cutoff frequency of the proposed filter is consistent with that of the 2.5-D SSPP TL. The gray part represents the dispersion rejection band, and the vertical dotted line is the cut-off frequency of mode 1. It can be seen that the dispersion rejection band is located in the rejection band of the filter and the cut-off frequency is consistent, which verifies the consistence between the dispersion characteristics and the transmission features.

To analyze the influence of the parameters of the proposed 2.5-D SSPP structure on filtering features, a parametric study is performed. From Fig.3(a)(b)(c), it can be seen that as adjusting the values of $g$, $w_b$ and $l_b$ can significantly change the rejection band center frequency while slightly changing the rejection band bandwidth and keeping the cut-off frequency of mode 1 almost unchanged. While, as is shown in Fig.3(d), changing the value of $d_t$ can effectively control the cut-off frequency of mode 1 without affecting the rejection band performance.

Therefore, the rejection band and the cut-off frequency of the rejection filter can be tuned independently by properly adjusting the geometrical parameters of the 2.5-D SSPP structure and the folded SRRs.

### C. Measurement results

The photo as well as the simulated and measured S-parameters of the fabricated rejection filter are shown in Fig. 4. The measured IL is less than 2.0 dB and the measured return loss is less than 10 dB within both passbands. The rejection band locates around 6.13 GHz with its 3-dB bandwidth from 5.92 to 6.32 GHz, and the maximum rejection depth is up to -40 dB. The measurement results agree well with the simulation results and slight differences between them are caused by fabrication and measurement errors.

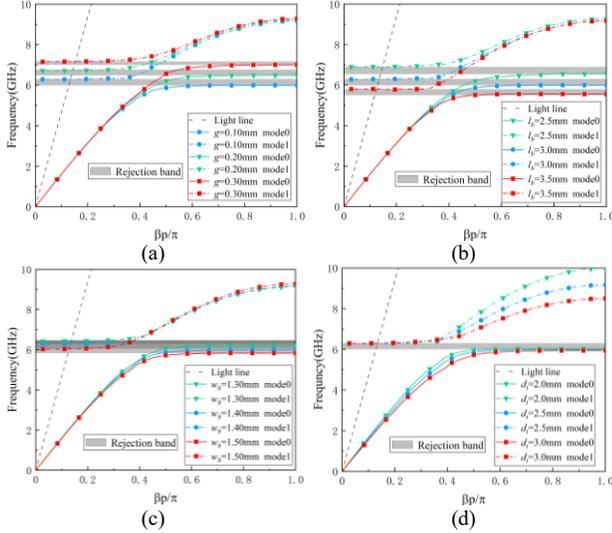

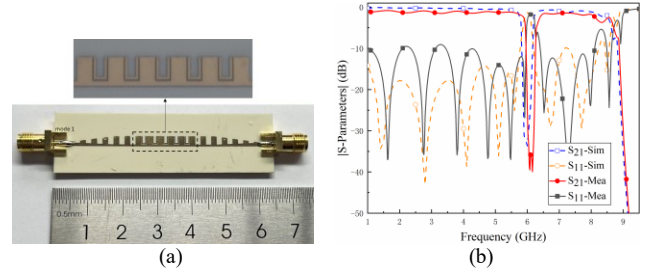

Fig. 4. (a) Photo of the fabricated rejection filter. (b) Simulated and measured |S-parameters| of the proposed rejection filter.

Fig. 3. The S-Parameters of the proposed rejection filter with different (a) $g$, (b) $l_b$, (c) $w_b$, and (d) $d_t$.

TABLE II
PERFORMANCE COMPARISONS OF SSPP REJECTION FILTERS

| Ref. | Lateral Size[1] | Num. | IL (dB) | $f_0$ (GHz) [2] | Rejection Depth (dB) |
|---|---|---|---|---|---|
| [5] | $1.40\lambda_g$ | 8 | 4.6 | 6.75 | -15 |
| [6] | $3.41\lambda_g$ | 6 | 3.1 | 8.74 | -33 |
| [7] | $0.15\lambda_g$ | 5 | 4.3 | 13.5 | -43 |
| [8] | $0.24\lambda_g$ | 5 | 2.5 | 18.43 | -42 |
| This work | $0.11\lambda_g$ | 5 | 2.0 | 6.13 | -40 |

[1] $\lambda_g$ is the guided wavelength of the central rejection frequency.
[2] $f_0$ is the central rejection frequency.

Table II compares the lateral size and number of the rejection units, insertion loss, and rejection depth of the proposed 2.5-D SSPP rejection filter with those of previous SSPP rejection filters [5-8]. It is shown that the proposed structure has significantly decreased the lateral size. Compared with that of the 2-D SSPP rejection filter [8], the lateral size of the proposed 2.5-D SSPP rejection filter is reduced by 54%. It can also be seen that the proposed structure has insertion loss of 2.0 dB and good rejection performance, which is achieved by using small number of rejection units.

## IV. CONCLUSION

In this paper, a compact rejection filter based on 2.5-D SSPP and folded SRRs is proposed. Compact folding structure results in the lateral size reduction by 54% compared with the conventional 2-D SSPP rejection filter. The proposed filter generates a rejection band with rejection depth up to -40 dB in the ultra-wide passband, while maintaining low passband insertion loss of 2.0 dB. Therefore, it is suitable for the application in highly-integrated microwave systems.

ACKNOWLEDGEMENT

This work was supported by the National Key Research and Development Program of China No. 2019YFB1802904.